\documentclass{vgtc}                          % final (conference style)

\ifpdf%                                % if we use pdflatex
  \pdfoutput=1\relax                   % create PDFs from pdfLaTeX
  \usepackage{graphicx}                % allow us to embed graphics files
  \DeclareGraphicsExtensions{.pdf,.png,.jpg,.jpeg} % for pdflatex we expect .pdf, .png, or .jpg files
\else%                                 % else we use pure latex
  \usepackage{graphicx}                % allow us to embed graphics files
  \DeclareGraphicsExtensions{.eps}     % for pure latex we expect eps files
\fi%

\usepackage{microtype}                 % use micro-typography (slightly more compact, better to read)
\PassOptionsToPackage{warn}{textcomp}  % to address font issues with \textrightarrow
\usepackage{textcomp}                  % use better special symbols
\usepackage{mathptmx}                  % use matching math font
\usepackage{times}                     % we use Times as the main font
\usepackage[T1]{fontenc}
\usepackage{cite}                      % needed to automatically sort the references
\usepackage{tabu}                      % only used for the table example
\usepackage{booktabs}                  % only used for the table example

\onlineid{1059}

\vgtccategory{Research}

\vgtcinsertpkg

\title{A Literature-based Visualization Task Taxonomy for Gantt Charts}

\author{Sayef Azad Sakin\thanks{email: sayefsakin@sci.utah.edu}%
\and Katherine E. Isaacs\thanks{e-mail: kisaacs@sci.utah.edu}}%
\affiliation{\scriptsize SCI Institute and Kahlert School of Computing, The University of Utah}

\abstract{
Gantt charts are a widely-used idiom for visualizing temporal discrete event sequence data where dependencies exist between events. They are popular in domains such as manufacturing and computing for their intuitive layout of such data. However, these domains frequently generate data at scales which tax both the visual representation and the ability to render it at interactive speeds. To aid visualization developers who use Gantt charts in these situations, we develop a task taxonomy of low level visualization tasks supported by Gantt charts and connect them to the data queries needed to support them. Our taxonomy is derived through a literature survey of visualizations using Gantt charts over the past 30 years.

}


\begin{document}

\firstsection{Introduction}

\maketitle
Gantt charts are a visual idiom for displaying multiple event sequences with dependencies between events in different sequences. Data is organized on two axes, one representing a notion of time or order and the other partitioning the events by some factor relating them, such as different people, roles, or resources assigned to them.

Gantt charts are widely used in project planning, process scheduling, and progress tracking~\cite{wilson2003gantt}. Project managers use Gantt charts for planning employee work assignments and estimating project completion time~\cite{tory2013comparative}. Gantt charts are similarly used in manufacturing for scheduling multi-stage pipelines~\cite{jo2014livegantt}.

Medical practitioners have used Gantt charts to understand treatment responses across individuals~\cite{antweiler2022uncovering}. In parallel computing, Gantt charts are used to examine inter-resource dependencies and complex program behavior~\cite{isaacs2014state}.

However, many of these applications, especially those coming from automated processes such as manufacturing and computing, can generate data with billions of events across tens of thousands of independent sequences---a scale beyond most Gantt chart implementations. A variety of designs have been proposed for when the data can no longer be resolved in a single image. Often the strategy involves aggregating or eliding data and implementing interactions such as zooming, panning, and filtering to present (a subset of) the data in the intuitive Gantt chart form. However, poor support for scale continues to be a significant barrier~\cite{isaacs2014state, davidson2023qualitative} and querying these large-scale datasets to implement the existing techniques can lead to large latencies, further hampering exploratory analysis~\cite{liu2014latency}. 

To better understand strategies and barriers to using Gantt charts with large numbers of events or sequences, we develop a literature-based task taxonomy, focusing on low-level visualization tasks that Gantt-focused systems use to support scale. Our taxonomy includes not only the Gantt-based tasks, but also those of highly coupled auxiliary views supporting the analysis of Gantt data. This taxonomy can be matched with user tasks when designing with Gantt charts in single or multi-view systems. We also match these tasks with data queries supporting them, enabling the analysis of data management techniques for supporting interactions as the datasets scale up. We expect visualization creators to use this taxonomy when considering options for representing large scale Gantt data and when considering implementation needs.

In summary, our contributions are:

\vspace{-0.5ex}

\begin{itemize}
    \itemsep=-0.5ex
    
    \item A multi-layer \textbf{visualization task taxonomy} for interactive Gantt charts and their common auxiliary views,
    
    \item A \textbf{classification of data queries} and their relationship with the proposed visualization task taxonomy for representing discrete event data visualized by Gantt charts, and
    
    \item  A \textbf{literature review} of scholarly articles over the last 30 years that have used Gantt charts as significant view.
\end{itemize}

\section{Gantt Chart Terminology and Related Work}
\label{sec:background}

We discuss Gantt chart data, encoding, and related surveys. 
\begin{figure}[ht!]
  \centering
  \includegraphics[width=1.0\columnwidth]{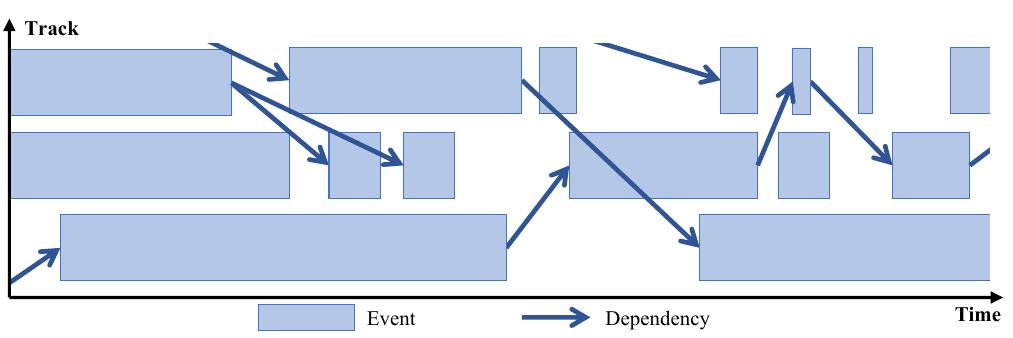}
  \caption{Gantt chart showing a window of time and three tracks.}
  \label{fig:gantt_desc}
\end{figure}

\textbf{Data.} The data items represented by Gantt charts are discrete {\em events} that have a partial order, typically due to some form of temporal attribute. We will refer to this ordering attribute as {\em time}. Thus, Gantt chart data fall into the general category of {\em event sequence data}. Gantt charts are typically used when the events have a {\em duration}, i.e., the start and end times are different. 

The events are typically divisible into multiple sequences by one or more of their attributes, e.g., events performed by different people can be organized as per-person sequences. 

The literature has used several terms for the attribute used to divide sequences, such as schedule, location, resource, person, process, or task. We choose the term {\em track} as it is still general but less overloaded in comparison to others (e.g., location, schedule, task, process). While time has an intrinsic order, tracks may have zero or more meaningful orderings. 

Gantt charts are commonly used when there are relationships between events of different tracks. For example, in parallel computing, processor 1 might execute a function (event) only after receiving a message created by a function (event) on processor 2. We refer to these kinds of cross-track relationships as {\em dependencies} and the events they connect as {\em dependent events}. The set of dependencies and dependent events together forms a {\em dependency graph}.

Events may have several more attributes associated with them in addition to the temporal, track, and dependency attributes. Some datasets may have multiple attributes that could be considered for time or track. While the track is an event attribute, it may also be treated as the central data unit for another view. For example, if tracks are employees, ``employees'' could be the data item of focus for another visualization. Thus, tracks can have additional attributes associated with them.

\textbf{Encoding.} The most prevalent form of a Gantt chart (\autoref{fig:gantt_desc}) is a two-dimensional visualization where time is mapped to the horizontal axis and tracks are displayed on the vertical axis. Each event is represented as a rectangular bar drawn according to its start time, end time, and track along these axes. Thus, the length of the bar represents its duration. Interactive Gantt charts allow panning and zooming on both axes.

Other arrangements exist, such as swapping the horizontal and vertical axes~\cite{de2010zinsight} or creating three-dimensional Gantt charts with a two-dimensional track layout, sometimes for symmetry (e.g., a circular layout~\cite{devaux2014datatube4log}) or geographical~\cite{schnorr2010triva} relations between tracks. 

We focus on the two dimensional form only in this work.

Dependencies between events are encoded as line segments or arrows connecting their associated dependent events.

Our examples use straight lines for dependencies, but orthogonal edges are sometimes used, especially in commercial project management tools~\cite{aigner2011visualization}.

While the organization of bars is the most salient feature of a basic Gantt chart, frequently auxiliary encodings focus on the dependencies. By making the dependencies and dependent events more salient, the displays becomes more like a node-link diagram of the dependency graph. 

Using a dependency graph to help manage the scale and complexity of the Gantt chart is a common strategy found in our literature review and thus we included it in our analyses.

\textbf{Related Work.} Several works~\cite{aigner2011visualization, shurkhovetskyy2018data, guo2021survey} survey temporal or event sequence data broadly, providing a general overview of time-oriented data visualization techniques.  

Parallel computing-specific surveys~\cite{ezzati2017multi, isaacs2014state} demonstrate Gantt charts for large-scale trace data among others. Peiris et al.~\cite{peiris2022data} developed a task taxonomy for timestamped instantaneous events without dependencies. Our work goes further in depth with Gantt charts specifically.

\section{Gantt Chart Task Taxonomy and Data Queries}
\label{sec:task_taxonomy}

\begin{figure*}[ht!]
  \centering
  \includegraphics[width=\textwidth]{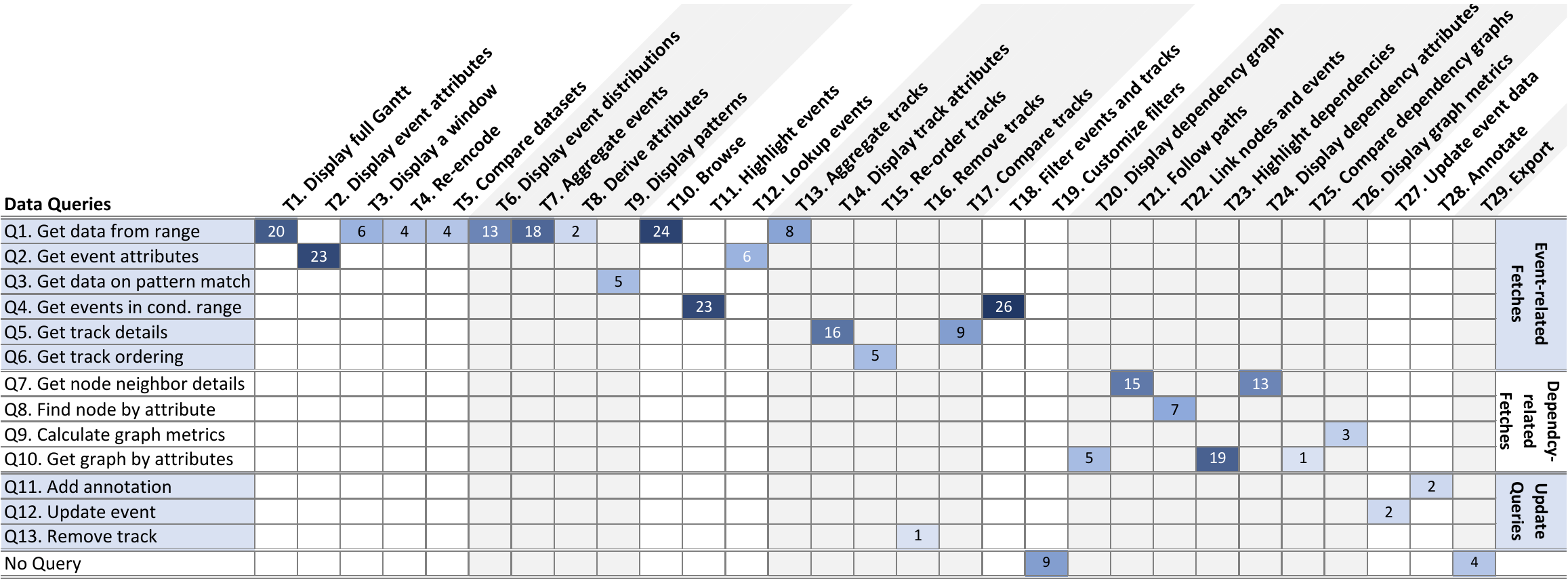}
  \caption{Mapping between data queries and interactive visualization tasks for Gantt charts. Each heatmap cell denotes in how many papers we observed the given visualization task. Tasks observed in each paper are supplemental materials. These counts suggest common tasks to support and possibilities for handling scale. The counts indicate which queries are the most common and may be considered for optimization.}
  \label{fig:gantt_heatmap}
\end{figure*}

We conduct a survey of Gantt charts used in visualizations with an emphasis on the visual tasks supported. We present our methodology, the derived tasks, and a translation of the tasks discovered into a set of data queries. \autoref{fig:gantt_heatmap} presents a heatmap of our visualization tasks and their mapping to data queries where each cell indicates in how many papers the task was observed.

\subsection{Methodology}
\label{subsec:task_methods}

We obtained an initial list of papers from surveys that feature Gantt charts~\cite{isaacs2014state, guo2021survey, ezzati2017multi}. We then expanded our list through searches in IEEE Xplore and Google Scholar using the following keywords, sourced from the surveys: \textit{Gantt chart},  
\textit{timeline visualization},  \textit{event sequence visualization}, \textit{schedule visualization}, \textit{execution trace visualization}, \textit{trace analysis}, \textit{performance analysis}, \textit{performance visualization}, \textit{parallel program analysis}, \textit{task parallel trace analysis},  and \textit{time-series charts}. We collected papers with promising abstracts, resulting in 137 research articles. We filtered these papers to those including interactive Gantt charts, resulting in a set of 35 papers. 

We also identified 11 additional commercially available online tools that use Gantt charts (see supplemental materials). We did not find additional tasks by investigating their websites. We do not include them in \autoref{fig:gantt_heatmap} as their Gantt view features were not described in detail in their publicly available documentation.

For each paper, we
examined at the description of use, task analyses if they existed, the visualization itself, and any use cases or evaluation. The initial round of qualitative coding focused on identifying, distilling, and tagging how the Gantt chart was used---what visualization tasks it was supporting. We then made a second pass to merge, split, and elaborate on the codes. This two-round process resulted in 29 distinct low-level visualization tasks. We categorized them into groups based on their intended higher-level user goal.

\subsection{Gantt chart task taxonomy}
\label{subsec:tasks}

We define the 29 visualization tasks, T1 - T29, collected into eight groups by their higher level user goals. We cite example sources from the literature leading to each task. The full list of sources for each task are presented in the supplemental material. 

These are {\em visualization} tasks, focused on the functionality the Gantt chart and its closely related auxiliary views need to provide over what the user is trying to accomplish. Choosing this perspective supports both the evaluation of data management strategies and tasks to support when implementing Gantt charts.

\subsubsection{Event Overviews}
These tasks involve viewing some or all of the discrete event data, often to get an overview or general context.

\textbf{T1: Display the full Gantt chart.} This is the most basic draw operation of the Gantt chart. For all but the smallest datasets, rendering the chart to fit on screen will involve some sort of aggregation, filtering, or overplotting, but we assume in our data queries that all of the data must be accessed. Often, visualization designers use a highly aggregated view as an overview to help with further navigation~\cite{nagel1996vampir, zaki1999toward, pinto2016analyzing, kaldor2017canopy}. 

\textbf{T2: Display event attributes.} Basic Gantt charts only directly encode time and track attributes. Color and labels are sometimes used to show additional attributes or auxiliary views may be used. Gantt charts often support requesting additional attributes interactively, initiated by brushing over the chart~\cite{andre2007continuum, meyer2016visual} or by clicking~\cite{pillet1995paraver} or hovering~\cite{de2010zinsight} on individual bars. The attributes are then shown in a separate static view~\cite{meyer2016visual, sakin2022traveler, kaldor2017canopy}, multiple coordinated views~\cite{nagel1996vampir,de2010zinsight,sakin2022traveler}, via lenses~\cite{nagel1996vampir, luz2010improving, jo2014livegantt}, or tooltips~\cite{de2010zinsight, jo2014livegantt, antweiler2022uncovering}.

\textbf{T3. Display a window.} This task shows a subset of the data in time and/or track space. It can be an outcome of other tasks (e.g., T10 - Browsing) or of directly specifying the window. Common modes of access include brushing over time~\cite{adhianto2010hpctoolkit, sakin2022traveler}, zooming, panning, textual input, or selecting through a derived dependency diagram~\cite{de2010zinsight}. 

\textbf{T4. Re-encode.} This tasks describes alterations of the Gantt encoding, such as mapping an axis to a different attribute set~\cite{pillet1995paraver, gupta2016movementslicer} or changing the domain of an existing axis, such as switching the notions or calculation of time~\cite{isaacs2014combing, jo2014livegantt, zaki1999toward}.

\textbf{T5. Compare events from multiple datasets.} This task presents multiple datasets. This is presented either with an additional window~\cite{sakin2022traveler, kaldor2017canopy} or within the same Gantt window~\cite{bell2003paraprof}.

\subsubsection{Analyze Derived Data}
Users may wish to explore derived data, such as summaries, aggregations, or clusters. These can help users understand their data at a higher level than individual events.

\textbf{T6. Display event distributions.} This task provides context to event attributes. Common methods include showing percentages of the whole~\cite{adhianto2010hpctoolkit}, histograms~\cite{sakin2022traveler}, or outliers to some condition~\cite{jo2014livegantt, schaubschlaeger2003event}. 

\textbf{T7. Aggregate events.} This task combines multiple events into a derived value such as the average, minimum, and maximum of certain attribute values, often time. Gantt chart-based visualizations have displayed the aggregated events several ways, including with a separate aggregate Gantt view~\cite{sakin2022traveler, nesi2023summarizing, kaldor2017canopy}, a magnification lens~\cite{jo2014livegantt}, re-encoding the chart area~\cite{topol1998pvanim}, and using juxtaposed marks~\cite{gupta2016movementslicer}.

\textbf{T8. Derive event attributes from dependencies.} Some analyses involve calculating metrics based on dependencies. For example, in computing contexts, `inclusive time' can be computed by summing durations of dependencies of an event. These derived metrics may be presented within the Gantt chart or separately~\cite{adhianto2010hpctoolkit, sakin2022traveler}.

\textbf{T9. Display patterns.} Patterns of events and dependencies that can be defined, identified, and displayed~\cite{de2010zinsight, isaacs2014combing} for some data. 

\subsubsection{Search}  

These tasks support various forms of searching.

\textbf{T10. Browsing.} This task encompasses zooming and panning in the Gantt chart, leading to T3 (displaying a window). Most of the visualizations in our literature review support this task.

\textbf{T11. Highlight events.} Highlighting events based on some attribute value is prevalent in Gantt chart visualization and is essentially a locate-type user task. It is complementary to event filtering (T18). Commonly, events are highlighted with color~\cite{adhianto2010hpctoolkit, isaacs2014combing, zaki1999toward, sakin2022traveler, nesi2023summarizing, kaldor2017canopy, dkabrowski2024manufacturing}, but other encodings, such as additional visual marks~\cite{sun2021daisen} or re-organizing the Gantt bars~\cite{jo2014livegantt,luz2010improving}, have also been used.

\textbf{T12. Lookup events.} This task involves displaying the context of a particular event within the Gantt chart and/or auxiliary views. It is often the outcome of selection in one view causing related views to update. For example, 

clicking on a line of code could re-center the Gantt chart to the corresponding bar. Alternatively, clicking on a Gantt bar could highlight source code~\cite{yan1995performance, adhianto2010hpctoolkit, zaki1999toward} or related geographic data in a map~\cite{gupta2016movementslicer} (e.g., where the event occurred).

\subsubsection{Manipulating Tracks}
These tasks involve manipulating tracks in some way, typically interactively. Users may want to manipulate tracks to re-organize the data, simplify the view, or do comparisons.

\textbf{T13. Aggregate tracks.} This task combines multiple tracks, often requiring aggregating the events among them. The task can be pre-computed~\cite{sakin2022traveler, sun2021daisen, nesi2023summarizing} or performed interactively~\cite{plaisant1998information, jo2014livegantt, kaldor2017canopy}.

\textbf{T14. Display track detail.} This task helps users understand how a track attributes may relate to event data. Gantt charts have supported showing track details through tooltips~\cite{sun2021daisen}, track-focused lenses~\cite{jo2014livegantt}, color encoding~\cite{schnorr2010triva}, highlighting~\cite{drebes2014aftermath, adhianto2010hpctoolkit}, and drawing additional rows to contain the detailed data~\cite{plaisant1998information}. 

\textbf{T15. Re-order tracks.} Users may want to change the order in which tracks are presented, manually or by attribute. Implementations include draggable rows~\cite{de2000paje}, re-ordering by a specific event attribute~\cite{heath1991visualizing}, and inputting a re-ordering condition through a separate interface~\cite{jo2014livegantt, de2010zinsight}. 

\textbf{T16. Remove tracks.} This task enables users to completely remove individual tracks from the chart. Of our surveyed papers, only Gupta et al.~\cite{gupta2016movementslicer} support this task. 

\textbf{T17. Compare tracks.} Though users can compare tracks if they are all already on screen, additional measures can be taken to ease visual comparison. 
 Techniques like repositioning tracks~\cite{sun2021daisen} or hiding unrelated tracks~\cite{koike1997visualinda,tory2013comparative, kaldor2017canopy} ease direct comparison. Pinto et al.~\cite{pinto2016analyzing} added percent usage summaries to each track to support aggregate comparison.

\subsubsection{Filtering}
These tasks support users focusing on specific events.

\textbf{T18. Filter events and tracks.} This task enables filtering events and/or tracks depending on event attributes. For example, in the medical domain this task could involve filtering patients (tracks) with a specific disease progression. In the computing domain, this task might be to filter down to events matching a certain library. This task was implemented by most of our surveyed papers with filtering in time being the most common.

\textbf{T19. Customize filters.} Complicated filters often require some sort of an interface to construct. Several works~\cite{fails2006visual, jo2014livegantt, drebes2014aftermath, pillet1995paraver, kaldor2017canopy} add a dedicated window with dropdown lists, checkboxes, and sliders to support building these filters. 

\subsubsection{Exploring dependencies} These tasks support analyzing dependencies. They may involve the dependencies as drawn in the Gantt chart or an auxiliary view.

\textbf{T20. Display dependency graph.} Rather than view only the Gantt chart which prioritizes events and tracks, users may want to focus on a graph (network) that can be constructed from the dependencies. Nodes in this graph can be events or, when shown in a separate view, aggregated groups of events. Displaying the graph can support identifying motifs in the graph that indicate some sort of behavior or pattern. In the literature, displaying the graph has been supported by de-emphasizing the other marks~\cite{schnorr2010triva} or using an additional view for the graph~\cite{adhianto2010hpctoolkit, de2010zinsight, sakin2022traveler}. 

\textbf{T21. Follow paths.} This task supports users in following chains of dependencies. It has been supported through Gantt chart navigation~\cite{haugen2015visualizing, hirakata2022exploring} or linked through the external graph view~\cite{sakin2022traveler, antweiler2022uncovering, kaldor2017canopy}. Interactive features, such as collapsing sub-graphs~\cite{de2010zinsight, sakin2022traveler, adhianto2010hpctoolkit} further help to navigate through a very large graph structure. 

\textbf{T22. Link nodes and events.} In the full dependency graph, nodes are events. However, manipulating the graph can lead to nodes that represent aggregated events. This task links the node back to its composite events, typically as a color highlight~\cite{de2010zinsight, sakin2022traveler} of both the graph nodes and the Gantt bars. 

\textbf{T23. Highlight dependencies.} This task draws a set of dependencies in the Gantt chart (and dependency graph if separate). Often, this selection is done with a specific path (relating to T21) or sub-graph~\cite{sakin2022traveler, schnorr2010triva, isaacs2014combing, haugen2015visualizing, tory2013comparative, dkabrowski2024manufacturing}. 

\textbf{T24. Display dependency attributes.} This task enables representing additional details regarding dependencies in the Gantt chart (and dependency graph if separate). This is often done by color encoding~\cite{koike1997visualinda, drebes2014aftermath} and additional visual marking~\cite{de2010zinsight, isaacs2014combing, antweiler2022uncovering, dkabrowski2024manufacturing}. 

\textbf{T25. Compare dependency graphs.} Multiple dependency graphs can arise when splitting a dependency graph or considering other event attributes as dependency sources. This task encompasses techniques supporting their comparison. Sun et al.~\cite{sun2021daisen} used a separate component view with treemaps to support this task.

\textbf{T26. Display graph metrics.} Statistics about the dependency graph, such as depth or total number of nodes and edges, can lead to insights about the data. Notions such as the distance between two given nodes or their common ancestor may also be sought. These values are generally shown in a separate detail view~\cite{adhianto2010hpctoolkit, schnorr2010triva, tory2013comparative}.

\subsubsection{Modifying the data} These tasks support users in altering the dataset, typically for annotation or to clean or fix the data.

\textbf{T27. Updating event data.} This task allows users to change event attribute data. This has been implemented by overwriting the data when a user re-positions an event~\cite{jo2014livegantt, bell2003paraprof}.

\textbf{T28. Annotate.} This task involves adding text and additional markings to the chart as annotations. Annotation data is separate from the base temporal event sequence data.
Multiple tools~\cite{drebes2014aftermath, gupta2016movementslicer} support this task.

\subsubsection{Exporting} 
User may want to save the visualization outside of the tool.

\textbf{T29. Export.} Several visualizations~\cite{drebes2014aftermath, zaki1999toward, pillet1995paraver, sakin2022traveler} provide a feature to export in a shareable format such as a PNG or SVG.

\subsection{Data queries for Gantt chart}
\label{subsec:task_queries}
 We organize the data queries required for the tasks based on the similarity of their data access profile. Note T19 (filter customization) and T29 (export) do not require fetching additional data. 

\subsubsection{Event-Related Fetch Queries}
These queries fetch data associated with events.

\textbf{Q1. Get data from a range.} The range may be multi-dimensional, range e.g., time and track. {\em Input:} A set of range tuples for each dimension being sub-selected. {\em Output:} A set of events, each with a start time, end time, and track value. {\em Supported tasks:} T1, T3, T4, T5, T6, T7, T8, T10, and T13.

\textbf{Q2. Get the attribute values of an event.} {\em Input:} The event identifier. {\em Output:} A list of attribute names and their values. {\em Supported tasks:} T2 and T12.

\textbf{Q3. Get data matching pattern.} This is a class of queries as the calculation to do the pattern match may be intensive and will differ by application and use case. {\em Input:} A search range similar to Q1 and the pattern to be matched. {\em Output:} A a set of events, each with a start time, end time, and track value. {\em Supported tasks:} T9.

\textbf{Q4. Get events from a range meeting conditions.} This query is similar to Q1 and Q3, but additional conditions must be met for the data to be included. {\em Input:} Ranges as in Q1 and conditions the data must satisfy. {\em Output: } A set of events that meet the condition, each with a start time, end time, and track value. {\em Supported tasks:} T11 and T18. 

\textbf{Q5. Get track details.} This query can be implemented in two ways, by specifying the tracks directly or by specifying events and seeking their associated tracks. {\em Input:} A list of track identifiers OR a list of event identifiers. {\em Output:} Attribute-value pairs for each track in the list OR the set of tracks associated with the input events. {\em Supported tasks:} T14 and T17. 

\textbf{Q6. Get track ordering.} {\em Input:} a list of tracks and a condition for their ordering, for example, a track attribute. {\em Output:} The ordered tracks list. {\em Supported tasks:} T15.

\subsubsection{Dependency-Related Fetch Queries}
These queries focus on fetching data associated with dependencies or their induced dependency graph. Queries are often performed on nodes, which may be individual or collections of events.

\textbf{Q7. Get neighbor details of a node.} 
{\em Input:} a node in the dependency graph. {\em Output:} all neighboring nodes and their attribute-value pairs. {\em Supported tasks:} T21 and T24.

\textbf{Q8. Find a node by attribute.} The search attribute is frequently a  dependency. {\em Input:} An attribute name and value. {\em Output:} Nodes matching that attribute value. {\em Supported tasks:} T22.

\textbf{Q9. Calculate graph metrics.} The exact query depends on the graph metric. {\em Input:} None. {\em Output:} Metrics of the dependency graph such as height and number of nodes. {\em Supported tasks:} T26. 

\textbf{Q10. Get graph by attributes.} This query generates a graph based on a set of attributes. These attributes can affect connectivity or membership (e.g., creating a subgraph based on a provided range). {\em Input:} List of attributes and attribute ranges. {\em Output:} The constructed dependency graph. {\em Supported tasks:} T20, T23, and T25.

\subsubsection{Update queries.} These queries update the data. They each return a boolean value indicating whether the update was successfully executed. 

\textbf{Q11. Add annotation.} {\em Input:} Time and track value indicating location and content (text) of the annotation. {\em Supported tasks:} T28. 

\textbf{Q12. Update event.} {\em Input:} The event identifier and attribute-value pair to be updated. {\em Supported tasks:} T27.

\textbf{Q13. Remove track.} This query marks a list of tracks for omission in subsequent queries. {\em Input:} A list of track identifiers. {\em Supported tasks:} T16.

\section{Discussion and Conclusion}
\label{sec:conclusion}

We presented a visualization task taxonomy for Gantt charts, an intuitive and popular visual idiom for displaying interdependent event sequences. Our motivation in creating this taxonomy is to support visualization design when using Gantt charts, especially in cases where large data in terms of number of events or tracks necessitate interactions or additional views beyond the core chart. This taxonomy serves as both a language for describing tasks and a collection of design strategies found in the literature.

To further aid both design and research with regards to Gantt charts, we paired our task taxonomy with an analysis of the data queries required to draw them. Careful consideration of both the scaling of the representation, i.e., through the visual and interaction design, and the scaling of the data management to support these queries, are important for designing and implementing effective visualizations to handle the demands of large-scale data produced in areas like manufacturing and computing. 

Despite the prevalence of domain applications that can generate billions of events~\cite{isaacs2014state}, we note that most visualizations in our survey did not reach that scale in their demonstration, with the largest shown datasets limited to millions of events. Thus, there may be  tasks not yet represented in the literature which thus do not appear in our taxonomy. Limitations in managing the data for the the common tasks may be hindering creation of additional strategies as using systems in the large scale regime is cumbersome. Our list of data queries can serve as a basis for choosing and developing data management strategies to scale up interactivity in Gantt charts, allowing for further research and observation into how people explore large-scale Gantt data.

\acknowledgments{
This work was supported by the Department of Energy under DE-SC0022044 and DE-SC0024635.}

\bibliographystyle{abbrv-doi}

\bibliography{main}
\end{document}

\maketitle

\section{Guide to supplemental materials}
This document contains the description of all provided supplemental materials along with their names. Here, the following materials are included in the accompanying \texttt{.zip} file.
\begin{itemize}
    \item \textbf{PapersSummary - Keywords.csv}: This CSV file contains the mapping between papers and the Gantt chart visualization tasks found in them. The code for each paper in the column \texttt{paper} of the CSV are mapped to their full citations in \autoref{sec:papers}.
\end{itemize}

\section{Surveyed Papers and their CSV Code}
\label{sec:papers}
We use a short code in ``PapersSummary - Keywords.csv'' when mapping tasks to each paper. We list the codes below along with their full citation.

\begin{enumerate}
\item HPCToolkit \cite{adhianto2010hpctoolkit},
\item Aftermath \cite{drebes2014aftermath},
\item Vampir \cite{nagel1996vampir},
\item Perf-EV \cite{pinto2016analyzing},
\item Jumpshot \cite{zaki1999toward},
\item Smart traces \cite{osmari2014visualization},
\item TAU \cite{xie2018performance},
\item Ravel \cite{isaacs2014combing},
\item VET \cite{haugen2015visualizing},
\item Paraprof \cite{bell2003paraprof},
\item Zinsight \cite{de2010zinsight},
\item AIMS \cite{yan1995performance},
\item Paje \cite{de2000paje},
\item Paraver \cite{pillet1995paraver},
\item Charm++ \cite{kale2006scaling},
\item Pvanim \cite{topol1998pvanim},
\item Trivia \cite{schnorr2010triva},
\item Visualinda \cite{koike1997visualinda},
\item DeWiz \cite{schaubschlaeger2003event},
\item LiveGantt \cite{jo2014livegantt},
\item EW \cite{tang2021visualization},
\item Ep-Gantt \cite{antweiler2022uncovering},
\item HP \cite{fails2006visual},
\item LaifLines \cite{plaisant1998information},
\item Luz \cite{luz2010improving},
\item MovementSlicer \cite{gupta2016movementslicer},
\item TASM \cite{tory2013comparative},
\item Daisen \cite{sun2021daisen},
\item GH \cite{hirakata2022exploring},
\item ParaGraph \cite{heath1991visualizing},
\item Traveler \cite{sakin2022traveler},
\item PDP \cite{nesi2023summarizing},
\item Canopy \cite{kaldor2017canopy},
\item LogiX \cite{dkabrowski2024manufacturing}, and
\item JobShopWeb \cite{loch431proposal}.
\end{enumerate}

\section{Commercial tools using Gantt chart}
\label{sec:ctools}
In addition to the tools associated with the papers above, we also visited the sites of the following commercial tools found in our searches. We manually searched each site for additional tasks not represented in our literature review, but did not find any. Most sites do not describe in detail the Gantt chart-centered features of these tools, so we did not include them in our accounting of how many times each task was represented.

\begin{enumerate}
    \item Perfetto \cite{Perfetto}
    \item Jaeger \cite{Jaeger}
    \item Zipkin \cite{Zipkin}
    \item AWS X-Ray \cite{AmazonXRay}
    \item Datadog APM \cite{Datadog}
    \item Lightstep \cite{Lightstep}
    \item Flurry \cite{Flurry}
    \item New Relic \cite{NewRelic}
    \item Honeycomb \cite{Honeycomb}
    \item Elastic \cite{Elastic}
    \item Tableau \cite{Tableau}
\end{enumerate}

\bibliographystyle{abbrv-doi}

\bibliography{main}

\appendix